\documentstyle[editedvolume,epsfig]{crckapb} 
\newcommand{\apj}{\,{\it Ap. J.}}  
\newcommand{\apjs}{\,{\it Ap. J. Suppl.}}
\newcommand{\grad}{\vec{\nabla}} 
\newcommand{\ddt}[1]{\frac{\partial #1}{\partial t}} 
\newcommand{\DDt}[1]{\frac{d #1}{dt}} 
\newcommand{\ddy}[1]{\frac{\partial #1}{\partial y}} 
\newcommand{\gram}{\,{\rm g}} 
\newcommand{\secd}{\,{\rm s}} 
\newcommand{\yr}{\,{\rm yr}} 
\newcommand{\cm}{\,{\rm cm}} 
\newcommand{\mdense}{\gram\cm^{-3}} 
\newcommand{\column}{\gram\cm^{-2}} 
\newcommand{\gauss}{\,{\rm G}} 
\newcommand{\mdot}{\dot{m}} 
\newcommand{\delab}{\nabla_{\rm\! ad}{}} 
\newcommand{\simless}{\mathbin{\lower 3pt\hbox
   {$\rlap{\raise 5pt\hbox{$\char'074$}}\mathchar"7218$}}}
\newcommand{\simgreat}{\mathbin{\lower 3pt\hbox
   {$\rlap{\raise 5pt\hbox{$\char'076$}}\mathchar"7218$}}}
\begin{opening}
\title{Thermonuclear Burning on Rapidly \protect \\ Accreting Neutron Stars}

\author{Lars Bildsten}
\institute{Department of Physics and Department of Astronomy\\
           366 LeConte Hall, University of California, Berkeley\\
           Berkeley, CA 94720}

\end{opening}

\runningtitle{Neutron Star Nuclear Burning}

\begin{document}

Neutron stars in mass-transferring binaries are accreting the hydrogen
and helium rich matter from the surfaces of their companions.  This
article explains the physics associated with how that material
eventually fuses to form heavier nuclei and the observations of the
time dependent phenomena (such as Type I X-ray bursts) associated with
the thermally unstable thermonuclear reactions.  The majority of the
article is pedagogical. Rather than just report from the computer
output, we will (when possible) explain why things happen the way they
do. This approach invariably comes at the expense of numerical
accuracy but has the distinct advantage of explaining how the outcome
depends on the composition of the accreting matter, the accretion rate
and the mass, radius and thermal state of the neutron star.  We also
introduce many new analytic relations that are convenient for
comparisons to both observations and computational results.

 After explaining nuclear burning for spherically symmetric accretion
onto neutron stars, we discuss the possibility of asymmetric
burning. In particular, we discuss some of the mysteries from {\it
EXOSAT} observations of Type I X-Ray bursts and how the solution to
these puzzles may lie in considering the lateral propagation of
nuclear burning fronts around the star. Fully understanding this
problem requires knowledge of parameters previously neglected such as
the distribution of fresh fuel on the star, the magnetic field
strength, and the stellar rotation. Recent {\it RXTE} observations of
bursters may finally tell us some of these parameters.

\section{Introduction}

Matter accreted onto a neutron star of mass $M$ and radius $R$
releases $GMm_p/R\approx \ 200 \ {\rm MeV}$ per nucleon, a value much
larger than that released from thermonuclear fusion (7 Mev per nucleon
for fusing hydrogen to helium). Our lack of precise knowledge of the
global accretion rate ($\dot M$) and $M/R$ means that the luminosity
from steady-state nuclear burning (i.e. at the same rate as it is
accreted, $L_n=E_{nuc} \dot M$, where $E_{nuc}$ is the nuclear energy
released per gram) is impossible to distinguish from the accretion
luminosity.  Time dependent nuclear burning is more easily observed
and is identified by having a time-averaged luminosity $L_n$.

Hansen \& Van Horn (1975) first calculated how the accreted
matter fuses to heavier elements in steady-state, and showed that
almost all models were subject to the ``thin-shell'' thermal
instability of Schwarzschild \& H\"arm (1965). Though they did not
investigate the time dependence of the unstable nuclear burning,
Hansen and Van Horn concluded by noting:
\begin{quote}
{\it  ``The wide range of time scales found -- ranging from
milliseconds to months, depending upon the stellar model -- suggests
the importance of a continued search for periodic or quasi-periodic
phenomena over a similar range of time scales in the compact X-ray
sources . . . .''} 
\end{quote}
Such phenomena were soon found with the discovery of recurrent,
quasi-periodic, Type I X-ray bursts from low accretion rate ($\dot
M\simless 10^{-9} \ M_\odot \ {\rm yr^{-1}}$) neutron stars (Grindlay
et al. 1976, Belian, Conner \& Evans 1976). The successful association
(Woosley \& Taam 1976, Maraschi \& Cavaliere 1977, Joss 1977, Lamb \&
Lamb 1978) of the thermal instabilities found by Hansen \& Van Horn
(1975) with the X-ray bursts made a nice picture of a recurrent cycle
that consists of fuel accumulation for several hours followed by a
thermonuclear runaway that burns the fuel in $\simless 10$
seconds. The observational quantity $\alpha$ (defined as the ratio of
the time-averaged accretion luminosity to the time-averaged burst
luminosity) was of the 
order the expected value (i.e. $\alpha=(GM/R)/E_{nuc}\approx 15-30$) for a thermonuclear origin of the
bursts. This has proven to be one important test for attributing any
time-dependent phenomenon to nuclear burning. As we discuss here, the
nature of the time dependence need not always be the simple
limit-cycle behavior of Type I X-ray bursts.

 We are focused solely on neutron stars accreting globally at rates in
excess of $10^{-10} \ M_\odot \ {\rm yr^{-1}}$, which is appropriate
for most bright, persistent Low Mass X-ray binaries (in particular the
``Z'' and ``Atoll'' sources of Hasinger \& van der Klis 1989). We
begin in \S 2 by explaining the physics of steady-state nuclear
burning and discussing the ``thin-shell'' thermal instability. Section
3 is a derivation of the ignition conditions and an explanation of the
accretion rate dependent regimes of unstable burning. The energetics,
recurrence times and durations of the bursts are discussed in \S 4.
We do not review much of the pre-RXTE observations of Type I X-Ray
bursts, as there are many reviews for that material (see Lewin, van
Paradijs and Taam, 1995 for the most recent summary of observations of
Type I X-ray bursts, including details of radius expansion bursts
which are not discussed here).  However, we do summarize (in \S 4.2)
what was learned from {\it EXOSAT} observations of bursters, as these
pointed to the possibility of asymmetric burning.
  
Section 5 opens up the discussion of asymmetric burning and presents
what is known about the speed of propagation of burning fronts on
neutron stars. It is often said that ``X-ray Pulsars Don't Burst'' and
``X-ray Bursters don't Pulse''. Section 6 discusses this dichotomy in
some detail and points out those accreting pulsars which might someday
violate this ``rule'', especially the recently discovered pulsar GRO
J1744-28 (Kouveliotou et al. 1996, Finger et al. 1996). We close in \S
7 by summarizing the recent RXTE results on coherent oscillations
during Type I X-ray bursts.

\section{Thermal Stability of Steady State Nuclear Burning} 

 Most neutron stars are accreting a mix of hydrogen and helium from a
relatively unevolved companion star. The standard case we consider
here has a hydrogen mass fraction of $X=0.7$. In this case, most of
the nuclear energy release is from fusion of the hydrogen to
helium. The very high temperatures reached ($T\simgreat 10^9 {\rm \
K}$) when this burning is unstable leads to production of elements
near the iron group.  In the steady-state burning case, the
temperatures in the hydrogen/helium burning region are not as high and
so the fusion to the iron group elements occurs at much larger depths.
It is also important to keep in mind that the CNO metallicity
($Z_{CNO}$) of the accreting material can vary by a few orders of
magnitude, depending on what stellar population the companion is from.

\subsection{Equations and Definitions} 

Most workers use the global accretion rate, $\dot M$, as the
fundamental parameter for discussion of the state of nuclear
burning. However, the plane-parallel nature of the neutron star
atmosphere means that the physics of thermal stability and nuclear
ignition actually depends on the accretion rate per unit area, $\dot
m$ (Fujimoto, Hanawa \& Miyaji 1981, hereafter FHM).  This parameter
determines the local behavior on the neutron star and need not be the
same everywhere.  The local Eddington rate is
\begin{equation} 
\dot m_{\rm Edd}={{2 m_p c}\over {(1+X)R \sigma_{\rm
Th}}}={1.5\times 10^5 \ {\rm g \ cm^{-2} \ s^{-1}}\over (1+X)}
\left({10 \ {\rm km} \over R}\right),
\end{equation} 
where $\sigma_{\rm Th}$ is the Thomson scattering cross-section, $m_p$
is the proton mass and $c$ is the speed of light. The ram pressure
within the accreted and settling atmosphere is negligible, so that
hydrostatic balance yields $P=gy$, where $dy=-\rho dz$ is the column
depth (in units of g cm$^{-2}$) and $g\approx GM/R^2$ (we neglect
general relativistic corrections throughout this article). The local
pressure scale height, $h=P/\rho g$, is always $\ll R$ and so $g$ is
nearly constant and the atmosphere is geometrically thin.  The flux is
carried by radiative transport and, for the accretion rates of
interest here, the main atmospheric opacity is Thomson scattering,
$\kappa=\kappa_{es}=\sigma_{\rm Th}(1+X)/2m_p$.  The flux leaving the
plane parallel atmosphere is then
\begin{equation}\label{eq:heat}
F=-{c\over 3\kappa \rho}{d\over dz} aT^4={c\over 3\kappa }
{d\over dy} aT^4,
\end{equation}
where $a$ is the radiation constant. For a constant flux, this
integrates to 
\begin{equation}\label{eq:flux}
T^4={{3\kappa P F}\over {acg}}, 
\end{equation}
at pressures large relative to the
photospheric value $P_{ph}\approx g/\kappa$ 
(i.e. the radiative zero solution, Schwarzschild 1958). 

  Steady accretion modifies the equations of particle continuity and
entropy for hydrostatic settling of the atmosphere (FHM). One
rewrites these equations in a coordinate system of fixed pressure. In
this case the accreted matter flows through the coordinates as it is
compressed by accretion of fresh material from above.  The continuity
equation for an element $i$ (with number density $n_i$) is
\begin{equation}
\ddt{n_i} + \grad \cdot  (n_i\vec{v}) = \sum r,
\end{equation}
where $r$ is the sum of particle creation and destruction processes.
For high accretion rates, there is not time for 
differential settling (i.e.
diffusion of one charged species relative to another) and all elements
co-move downward at the speed needed to satisfy mass continuity,
$v=\dot m/\rho$ (Wallace, Woosley \& Weaver 1982, Bildsten, Salpeter
\& Wasserman 1993). 
We then define a mass fraction $X_i\equiv
\rho_i/\rho= A_i m_p n_i/ \rho $ where $A_i$ is the baryon number of
species $i$ and expand the continuity equation to obtain
\begin{equation}\label{eq:contin}
\ddt{X_i} + \dot{m} \ddy{X_i} = \frac{A_i m_p\sum r}{\rho}. 
\end{equation}
The equation for the entropy is 
\begin{equation}
T\DDt{s} = -\frac{1}{\rho}{\grad \cdot \vec{F}} + \epsilon,
\end{equation}
where $\epsilon$ is the energy release rate from 
nuclear burning. We write the entropy as 
$Tds=C_p T(dT/T-\delab dP/P)$, 
where $\delab=d\ln T/d\ln P$ for an adiabatic change and 
$C_p$ is the specific heat at constant pressure. 
Then since
the temperature can depend on both time and pressure, we find 
\begin{equation}\label{eq:entrop}
\ddy{F} + \epsilon=C_p\left(\ddt{T}+\dot{m}\ddy{T}\right)-\frac{C_pT\mdot}{y}\delab.
\end{equation}
These are the equations that describe the hydrostatic evolution of the
neutron star atmosphere and are valid while the matter accumulates on
the neutron star. The thermonuclear instability can sometimes lead to
outflow and hydrodynamics, for which these equations cannot be used.

\subsection{The Upper Atmosphere Before Burning}

The neutron stars in low-mass X-ray binaries accrete via an accretion
disk formed in the Roche lobe overflow of the stellar companion. There
are still debates about the ``final plunge'' onto the neutron star
surface, with some advocating that there is a strong enough magnetic
field to control the final infall, while others prefer an accretion
disk boundary layer. The only known effect on the later nuclear
burning is that substantial destruction of all elements heavier than
helium via spallation reactions can occur if the final plunge has an
appreciable radial component (Bildsten, Salpeter \& Wasserman
1992). The predicted amounts of spallation are substantial and
basically turn the accreted matter into hydrogen, helium and a mix of
light fragments, none of them capable of sustaining a  CNO
cycle.

Eventually the accreted material reaches the stellar surface,
thermalizes and radiates away nearly all of the $\approx 200 \ {\rm
MeV\ nucleon^{-1}}$ infall energy. Hence, the flux exiting the star a
few scale heights beneath the thermalizing region is from energy
released much deeper within the star, due to gravitational settling
and nuclear burning.  At this point, the matter is part of the star
and is in hydrostatic balance. It is continuously compressed by the
accretion of new material from above and eventually reaches pressures
and temperatures adequate for thermonuclear ignition. The equations
describing the settling and compression of the the atmosphere are heat
transfer (equation [\ref{eq:heat}]) and entropy (equation
[\ref{eq:entrop}]). Rather than give a detailed discussion of these
``settling'' solutions, let's just compare terms in the entropy
equation in order to show what is most important.

The compression occurs on a timescale $t_{accr}\approx y/\dot m$,
which we compare to the thermal timescale set by radiative diffusion
\begin{equation}\label{eq:thermal}
t_{th}\approx {3\kappa C_p y^2\over 4acT^3},
\end{equation}
in order to discover whether the compression is adiabatic. Presuming
that the atmosphere at some column depth $y$ is carrying a flux, $F$,
then the ratio of these two timescales is
\begin{equation}
{t_{th}\over t_{accr}}\sim {C_p T \dot m\over F},
\end{equation} 
so that, when $F \gg C_p T \dot m\approx 5k_BT\dot m/2\mu m_p$ the
thermal time is much less than the rate of compression, where $\mu$ is
the mean molecular weight of the accreting gas. This is always the
case in the upper atmosphere ($\rho< 10^6 \ {\rm g \ cm^{-3}}$), as
the energy released per proton from either steady burning or deeper
gravitational settling is much more than $k_BT$ in the upper
atmosphere.  In this regime, the ``trajectory'' of the compressed
fluid element is far from adiabatic as it has time to transport heat
while it is being compressed.  This simplifies the calculation, as the
temperature gradient in the upper atmosphere is then set by the
flux. The outer boundary condition plays no role.\footnote{The problem
is much more subtle when the thermal time at some depth is comparable
to the local compression time, $y/\dot m$.  For neutron stars, this is only
known to happen at accretion rates far in excess of the Eddington rate
at depths beneath the hydrogen and helium burning (Bildsten \& Cutler
1995, Brown \& Bildsten 1997).}  These settling solutions are valid
until the helium burns fast enough to appreciably change either the
temperature or the helium abundance.

For the high accretion rates of interest here, most of the outer
envelope is non-degenerate so $P=\rho k_B T/\mu m_p=gy$. We can
always correct the equation of state for degeneracy by reducing
$\mu$ to $\mu_{eff}=\rho k_B T/Pm_p<\mu$. 
The temperature profile is determined by the flux
flowing through the atmosphere, which is presumed to be a constant
$F=E_{atm}\dot m$, where $E_{atm}$ is the energy released per accreted gram
at some large depth in the atmosphere. This parameter will be
determined later and depends on whether the burning is in steady-state
or not.  Then when the flux is written as $F=10^{23} E_{18}\dot m_5 \
{\rm ergs \ cm^{-2} \ s^{-1}}$, where $E_{atm}=E_{18}10^{18} \ {\rm
ergs \ g^{-1}}$ and $\dot m=\dot m_5 10^5 \ {\rm g \ cm^{-2} \
s^{-1}}$, the relation between the density and temperature in the
upper atmosphere from equation (\ref{eq:flux}) is
\begin{equation}\label{eq:trho} 
T_8=3.32\rho_5^{1/3}\left({E_{18}\dot m_5(1+X)\over g_{14}}{0.6\over \mu}\right)^{1/3}
\end{equation}
where $T=T_8 10^8 K$, $\rho=\rho_5 10^5 \ {\rm g \ cm^{-3}}$ and
$g=g_{14}10^{14} \ {\rm cm \ s^{-2}}$.\footnote{I use $\dot m_5$ as the
local unit for the accretion rate, which is 
related
to the global supply onto the neutron star via
$\dot M=\dot m_5 2\times 10^{-8} M_\odot \ {\rm yr^{-1}} (R/10 \ {\rm
km})^2$. } This gives a good
first cut at how the temperature changes with depth in the accreting
envelope above the burning region. 

As mentioned earlier, the short thermal time leads to an entropy loss
for the compressing fluid elements. Using the above relation, we find
that $s\propto \ln(T^{3/2}/\rho)\propto \ln(1/\rho^{1/2})$, and so it
is clear that the entropy of the accreting material actually decreases
as it is compressed to higher densities. This temperature profile is
convectively stable and will allow us to find where the nuclear
burning begins. The convectively stable atmosphere has internal
buoyancy that is measured by the Brunt-V\"ais\"al\"a frequency (Cox
1980) 
\begin{equation}
N^2=-g\left({d\ln \rho\over dz}-{1\over \Gamma_1}{d\ln P\over dz}\right)=
-g\left({d\ln \rho\over dz}+{1\over \Gamma_1 h}\right),
\end{equation}
where $h=P/\rho g\ll R $ is the local scale height and 
$\Gamma_1=(d\ln P/d\ln \rho)_{adiab}=5/3$ is the 
adiabatic index. The temperature-density profile for a 
constant flux atmosphere then gives
\begin{equation}\label{eq:brunt}
N^2={3g\over 20 h}={3\mu m_p g^2\over 20 k_B T}, 
\end{equation}
which decreases with depth in the star since the temperature rises.
The frequency of local buoyant oscillations is then $N\approx (50 {\rm
\ kHz} ) g_{14}/T_8^{1/2}$, comparable to the inverse of the time it
takes sound to travel a scale height.

\subsection{ How Hydrogen Burns at High Accretion Rates}

 For these high accretion rates, the atmospheric temperature is always
in excess of $10^7$ K, so that hydrogen burns via the CNO
cycle. However, at large densities and high temperatures, the
timescale for proton captures becomes shorter than the subsequent
$\beta$ decay lifetimes.

For example, consider the $^{14}$N(p,$\gamma$)$^{15}$O reaction, which
is one of the slower steps in the cycle. The timescale for the proton
capture is $t_{cap}=(n_p\langle\sigma v \rangle)^{-1}$, where
$\langle\sigma v \rangle$ is the thermally averaged reaction rate from
Caughlan \& Fowler (1988), and becomes shorter than a typical $\beta$
decay time of 100 seconds when $T>8\times 10^7 \ {\rm K}$ for the
relevant density of $\rho=10^5\ {\rm g \ cm^{-3}}$. These temperatures
are usually reached before the unstable ignition of helium takes
place. The rapid proton captures makes a slightly different 
burning cycle than typically encountered, which is called the
hot CNO cycle (Hoyle \& Fowler 1965)
\begin{equation} 
^{12}{\rm C}(p,\gamma)^{13}{\rm N}(p,\gamma)^{14}{\rm
O}(\beta^+)^{14}{\rm N}(p,\gamma)^{15}{\rm O}(\beta^+)^{15} {\rm
N}(p,\alpha)^{12} {\rm C}  \end{equation} The time to go around this
catalytic hydrogen burning loop is set by the $\beta$ decay lifetimes
of $^{14}{\rm O}$ ($t_{1/2}=71$ s) and $^{15}{\rm O}$ ($t_{1/2}=122$
s) and is {\it temperature independent} and thermally stable.  While
this cycle is operative, all of the seed nuclei are locked up in
$^{14}{\rm O}$ and $^{15}{\rm O}$. These $\beta$-decay limitations fix
the hydrogen burning rate at
\begin{equation}
\epsilon_h=5.8\times 10^{15} Z_{\rm CNO} {\rm \ ergs  \ g^{-1} \
s^{-1}},
\end{equation} 
where $Z_{\rm CNO}$ is the mass fraction of CNO in the accumulating
matter.

There is thus a minimum column density of fresh fuel needed on the
star, $y_{h}$, to burn the hydrogen at the rate at which it is
accreted (i.e. steady-state).  It is found by setting $\epsilon_h
y_{h}=\dot m X E_{h}$ where $E_h\approx 6.4\times 10^{18} \ {\rm ergs
\ g^{-1}}$ is the energy released from hydrogen burning to helium,
giving
\begin{equation}\label{eq:yh} 
y_{h}\approx 10^{10} \ {\rm g \ cm^{-2}} X \dot m_5 \left({0.01\over
Z_{\rm CNO}}\right).
\end{equation} 
This is equivalent to the amount of matter accreted onto the star in
the time it takes to go around the cycle enough times to consume all
of the hydrogen, $E_{h}/\epsilon_h\approx (10^3/Z_{\rm CNO}) {\rm s}$,
or about one day for typical metallicities. However, the matter will
reach high enough temperatures within a day of landing on the neutron
star so that the primordial helium can ignite. In this case helium
burning occurs in a hydrogen-rich environment, which enhances the
nuclear reaction chains and energy release (Lamb \& Lamb 1978; Taam \&
Picklum 1978, 1979; FHM; Taam 1982).

\subsection{When Does The Helium Start Burning?}

Now let's ask when the helium ignites and whether or not it is
thermally stable. The energy generation rate for helium burning to
carbon is (Hansen \& Kawaler 1994) 
\begin{equation}
\epsilon_{3\alpha}=5.3\times 10^{21}{\rho_5^2 Y^3\over
T_8^3}\exp\left({-44\over T_8}\right)  {\rm ergs \ g^{-1} \ s^{-1}}.
\end{equation} 
In a steady-state burning situation, the 
helium depletes at the depth where the lifetime to the
nuclear reaction equals the time it takes to cross a scale height
(Taam 1981, Fushiki \& Lamb 1987) 
\begin{equation}\label{eq:time}
{\dot m Y\over y}\approx {\epsilon_{3\alpha}\over E_{3\alpha}}, 
\end{equation}
where $E_{3\alpha}=5.84\times 10^{17} \ {\rm ergs \ g^{-1}}$ is the
energy release from $3\alpha \rightarrow ^{12}{\rm C}$. Even when 
hydrogen is present, this condition is the appropriate one,
as the carbon produced from helium burning increases the
number of seed nuclei for the hot CNO cycle and allows for more
rapid hydrogen consumption. 

The helium depletion condition (equation
[\ref{eq:time}]) is now combined 
with the temperature profile (see equation
[\ref{eq:trho}]; we leave the energy released per accreted
particle, $E_{18}$,  as a free parameter) to find the temperature at
which the helium is burned
\begin{equation}
\dot m_5^4=68.2\left[{{(Yg_{14} \mu)^2 T_8^7} \over {(1+X)^3E_{18}^3}}\right]
\exp\left({-44\over T_8}\right).
\end{equation} 
Solving this transcendental equation for the temperature also gives
the pressure for the helium burning, and most importantly, how these
quantities depend on the local accretion rate, stellar mass and
radius, and abundances. In order to simplify the transcendental,
expand the exponential about the temperature of $T_8=3.38$ so that
$\exp(-44/T_8)\approx 2.22\times 10^{-6}(T_8/3.38)^{13}$. 
Some
accuracy is compromised 
because of this approximation, but not much. The resulting
equation for the burning temperature is
\begin{eqnarray}\label{eq:tburn}
T_{burn}&=&3.43 
\times 10^8 {\rm K} \dot m_5^{1/5} {{(E_{18}+XE_{18})^{3/20}} \over 
(Yg_{14}\mu)^{1/10}}, \\ \nonumber
&=& 
{\rm Temperature \ at \ Helium \ Burning \ Location}. 
\end{eqnarray} 
The very strong temperature sensitivity of the helium burning makes
this temperature estimate fairly accurate. The column density at the
burning location is
\begin{equation}\label{eq:yburn}
y_{burn}={{5.2\times 10^7 \ {\rm g\ cm^{-2}}}\over
{\dot m_5^{1/5}(E_{18}Y g_{14} \mu (1+X))^{2/5}}}, 
\end{equation}
which is a bit trickier to compare to a real calculation that finds
the abundance profiles self-consistently. 

 For example, a full integration from Bildsten \& Brown (1997) for a
star with $g_{14}=1.87$, $X=0.7$, and $Y=0.3$ accreting at $\dot
m_5=1$ and burning in steady-state (so that $E_{18}=5.3$) found that
one-half of the accreted helium is burned at a location where
$T_8=5.38$, $y=8.3\times 10^7 \ {\rm g \ cm^{-2}}$ and $\rho=2\times
10^5 \ {\rm g \ cm^{-3}}$. The matter is non-degenerate and the
opacity is slightly less than Thomson scattering $\kappa\approx
0.4\kappa_{es}\approx 0.136 \ {\rm cm^2 \ g^{-1}}$. Our formulae
(equations [\ref{eq:tburn}] and [\ref{eq:yburn}]) give $T_8=5.32$ and
$y_{burn}=3.3\times 10^7 {\rm g \ cm^{-2}}$ for this case. As you can
see the temperature is in excellent agreement. Even when we correct
for the lower opacity, the estimated column is nearly a factor of 2
low relative to the numerical estimate.  This is because we have found
the place where the lifetime of an alpha particle is comparable to the
time since arrival on the star. To completely burn the matter in
steady-state requires a time longer than this and is roughly
$t_{burn}\approx {10^3 {\rm s}}/{\dot m_5^{6/5}}$ or $<$ hour at
accretion rates where the burning is stable.

We can now check the previous statement that the helium ignites before
the hydrogen has completely burned. This is true for high accretion
rates, as $y_{burn}< y_{h}$ (compare equations [\ref{eq:yh}] and
[\ref{eq:yburn}]) as long as $\dot m_5>0.016(0.01/Z_{CNO})^{5/6}$.
All steady-state solutions that we find at high accretion rates are
burning the helium in a hydrogen-rich environment.

\subsection{ Thermal Stability of Steady-State helium Burning} 

 Let's now understand whether these models are stable to a thermal
perturbation. It is best to start by writing down the energy
equation. Since we are considering a thermal perturbation on a
timescale much shorter than the time it takes for the matter to move a
scale height, all terms $\propto \dot m$ in equation (\ref{eq:entrop})
can be dropped, giving
\begin{equation}
C_p{{\partial T}\over {\partial t}}= {{\partial F}\over {\partial y}}
+\epsilon.
\end{equation} 
It is a good approximation to presume a constant pressure during the
thermal perturbations. The simplest linear stability analysis of the
steady-state model presumes just one zone (i.e. single pressure,
Fujimoto et al. 1981) and so we set
\begin{equation}\label{eq:epscool}
{\partial F\over \partial y} \approx - {acT^4\over 3\kappa y^2}\equiv 
\epsilon_{cool},
\end{equation} 
as the simple representation of the radiative cooling rate from the
atmosphere. The $\beta$-limited hydrogen burning is temperature
independent, so that the instability must arise from helium
ignition. Writing $\epsilon_{3\alpha}\propto T^\nu\rho^\eta$, where
$\nu=44/T_8 -3 $, a constant pressure perturbation of the equilibrium
model (i.e. $\epsilon=-\epsilon_{cool}$) is stable when
\begin{equation} 
\nu-4+{{\rm
dln}\kappa\over {\rm dln} T}+{{\rm dln}\rho\over {\rm dln}
T}\left(\eta+{{\rm dln}\kappa\over {\rm dln} \rho}\right) < 0,
\end{equation}
where all the derivatives are at constant pressure.  These
steady-state solutions are only marginally degenerate (the density at
a given pressure and temperature never differs from the non-degenerate
guess by more than $15\%$) in the burning region and so ${\rm
dln}\rho/ {\rm dln} T=-1$.  In addition, the opacity is not far from
Thomson scattering, so those derivatives are zero, making the
condition just $\nu< 4+\eta$ for stability, clearly displaying that
the thermal stability is a competition between nuclear heating and
radiative cooling (which scales $\propto T^4$). We thus need 
$T> 4.88 \times 10^8 \ {\rm K}$ for 
stable helium burning. More sophisticated non-local linear analysis have
been carried out by Fushiki \& Lamb (1987).

For a thin shell, the degree of degeneracy has only a small effect on
the stability criterion and, most importantly, this instability does
not require that the matter be degenerate.  The more important
condition is that it is thin ($h\ll R$) so that it's temperature
increases when it is heated (i.e. a positive specific heat, $C_p>0$).
It is clearly thin before burning and remains so even during the
flash as the large gravitational well on the neutron star requires
temperatures of order $10^{12} \ {\rm K}$ for $h\sim R$. Temperatures this
high are never reached. 
This analysis provides a reasonable indicator of when we should expect time
dependence of the burning. The strength of the ``flashes'' is
sensitive to the non-linearity of the nuclear reaction rates, such as
the weakening of the temperature dependence at higher temperatures, and 
the decreasing density as the temperature rises at fixed pressure.

We are lucky that it is temperature which determines stability, as
this is the quantity that is best determined by this analytic
calculation. The above discussion basically pointed to the condition
$T_{burn}>4.88\times 10^8 {\rm K}$ for stability, which, from equation
(\ref{eq:tburn}) turns into a condition
\begin{equation}\label{eq:stabmdot}
\dot m > \dot m_{st}= 5.83 \times 10^5
{{\rm g}\over {\rm cm^2 \ s}}
{{(Yg_{14}\mu)^{1/2}}\over{[(1+X)E_{18}]^{3/4}}}
\left(\kappa_{es}\over \kappa\right)^{3/4} 
\end{equation}
for stable helium ignition. We first evaluate this
for accretion of pure helium and 
then discuss in more detail the more prevalent case of 
accretion of hydrogen-rich material. 
In both instances, we will define an accretion rate,
$\dot m_{st}$, above which the helium burning is thermally stable.

There are a few ultra-compact ($P_{\rm orb}<50 \ {\rm min}$) binaries
(4U~1820-30, 4U~1916-05 and 4U 1626-67, Nelson, Rappaport \& Joss
1986) which are most likely accreting pure helium from a degenerate
dwarf. The smaller nuclear energy released per gram for helium
accretion (relative to hydrogen-rich accretion) means that the flux
flowing through a steady-state burning envelope is less (Joss 1978,
Joss \& Li 1980, Bildsten 1995). Let's consider the steady-state
helium burning case, where $Y=1$, $X=0$, $\mu=4/3$ and $E_{18}=0.58$,
in which case the accretion rate needed for stability is
\begin{equation}
\dot m_{st}({\rm Pure \  Helium})\approx 
1.4\times 10^6 {{\rm g}\over {\rm cm^2 \ s}}
\left({M\over 1.4 M_\odot}\right)^{1/2}
\left({10 \ {\rm km} \over R}\right),
\end{equation}
20 times higher than the hydrogen-rich limit we will discuss
next. This is due to the reduced amount of energy release from helium
burning and allows for the bright helium accreting object 4U 1820-30
(see Bildsten 1995 for a detailed discussion) to be violently unstable
at accretion rates $\sim 10^{-8} M_\odot \ {\rm yr^{-1}}$. 

\subsection{ Stable Hydrogen and Helium Burning }

Most neutron stars are accreting from relatively unevolved companions
and so $X=0.7$, $Y=0.3$ and $\mu=0.6$. The energy released when
steadily burning all the hydrogen and helium to, say, carbon is
$E_{18}=5.3$. This gives a critical local accretion rate for stable
helium burning in this mixed environment of
\begin{equation}\label{eq:mdots} 
\dot m_{st}({\rm  H/He})\approx 
6.5\times 10^4 {{\rm g}\over {\rm cm^2 \ s}}
\left({M\over 1.4 M_\odot}\right)^{1/2}
\left({10 \ {\rm km} \over R}\right)
\left(\kappa_{es}\over \kappa\right)^{3/4}. 
\end{equation}
We can also write this in terms of the global Eddington rate
\begin{equation}\label{eq:mdotsted} 
{\dot M_{st}\over \dot M_{\rm Edd}}\approx 1.5\left({Y\over 0.3}{\mu\over
0.6}{M\over 1.4 M_\odot}\right)^{1/2} \left( 1+X\over 1.7\right)^{1/4}
\left(0.4 \kappa_{es}\over \kappa\right)^{3/4},
\end{equation}
where we have inserted the typical ratio of the opacities in the
burning region. There is no reason to believe that this calculation is
accurate enough to say whether the critical accretion rate for burning
is above or below the Eddington rate. What is more trustworthy are the
scalings, and notice that equation (\ref{eq:mdotsted}) is independent
of the stellar radius and weakly dependent on the stellar mass.  Using
the estimate of the critical accretion rate (equation
[\ref{eq:mdotsted}]), one finds that neutron stars brighter than
$L_{accr}=GM\dot M_{st}/R \approx 2\times 10^{38}(M/1.4 M_\odot)^{3/2}
\ {\rm ergs \ s^{-1}}$ should be stably burning the hydrogen and
helium. Note that a more massive neutron star can have unstable
nuclear burning at higher accretion luminosities than a less massive
star. This might provide a possible indicator of mass for those
objects where the distance is reasonably well known.

  Fujimoto et al. (1981) estimated $\dot M_{st}$ (they called it $\dot
M_{cri}$ in their Table 1) for 3 different neutron stars. They found
$\dot M_{st}/(10^{-9} M_\odot \ {\rm yr^{-1}})=15, \ 13, \ {\rm and} \
12$ for $M=0.476, 1 $ and $1.41M_\odot$. Our numbers inferred from
equation (\ref{eq:mdots}) are $\dot M_{st}/(10^{-9} M_\odot \ {\rm
yr^{-1}})=13, \ 16, \ {\rm and} \ 17$ for the same three masses and
radii. These are about $20-30\%$ different.  The most accurate way to
learn if the burning is thermally stable is to numerically solve the
time dependent equations (\ref{eq:flux}), (\ref{eq:contin}) and
(\ref{eq:entrop}). Ayasli \& Joss (1982) and Taam, Woosley \& Lamb
(1996) performed time-dependent calculations at accretion rates
comparable to the Eddington limit and found that the burning is still
thermally unstable when $\dot M\simless \dot M_{\rm Edd}$. Ayasli \&
Joss (1982) also carried out a few super-Eddington (by a few)
calculations and found thermal stability, in agreement with the
calculations here. However, there has not been a thorough study
(i.e. surveying the dependence of $\dot M_{st}$ on $M$, $R$ and
composition) of this problem. Indeed, little has been published on the
steady-state burning of matter on accreting neutron stars at
super-Eddington accretion rates.  The most likely environment for this
burning is on the polar caps of high accretion rate pulsars (see \S
6).

\section {Unstable Mixed Hydrogen/Helium Burning} 

Since most neutron stars in LMXB's are less luminous than $10^{38} \
{\rm ergs \ s^{-1}}$, we should expect some sort of time-variable
phenomena from nuclear burning.  We will find that, in one dimension,
the system evolves into a limit cycle, accumulating fuel for some long
time and then reaching a thermal instability which leads to observable
phenomena. 

\subsection{Thermonuclear Ignition Conditions } 

How much matter needs to be accreted before reaching an instability?
In order to keep things simple, let's presume that the accreted matter
has a metallicity large enough so that the flux leaving the accreting
envelope is just $F=\epsilon_{h} y_{t}$ prior to reaching an
instability, where $y_{t}$ is the column of freshly accreted matter on
the star since the last thermal instability (Taam 1980, Wallace,
Woosley \& Weaver 1982). \footnote{This requires that $\epsilon_h
y_t\simgreat k_BT/m_p \dot m$ so that we can neglect the energy
release from the compression of matter or just $Z_{CNO}\simgreat
10^{-3} T_8 \dot m_5/y_8$.  As we discuss in \S 3.3, this
condition will be violated for interesting values of the metallicity,
in which case the ignition conditions depend on the
accretion rate, $\dot m$.} The condition for the thermal instability
is slightly different for these non-equilibrium solutions, since the
entropy equation before the instability is balancing heat diffusion
with the energy released from the hot CNO cycle and gravitational
settling. However, neither of these energy sources can lead to a
thermal instability. It is the helium burning which takes care of
that.

The boundary of unstable/stable helium burning in the $y-T$
plane is defined by the relation 
\begin{equation}\label{eq:ignition}
{d\epsilon_{3\alpha}\over dT}=-{d\epsilon_{cool}\over dT},
\end{equation}
(FHM, Hanawa \& Fujimoto 1982, Fushiki \& Lamb 1987), where
$\epsilon_{cool}$ is defined by equation (\ref{eq:epscool}).  These
derivatives are taken at constant pressure, as the instability grows
faster than the pressure changes. Let's presume some knowledge about
the answer and expand the exponential in the helium burning rate at
$T_8=2.31$ so that we get a simple relation for the ignition
temperature on the unstable branch of the relation (equation
[\ref{eq:ignition}])
\begin{equation}\label{eq:helign}
T_{\rm ign}\approx {1.83\times 10^8 {\rm K}\over \kappa^{1/10} Y^{3/10}(\mu
g_{14}y_8^2)^{1/5}}. 
\end{equation}
The freshly accreted matter is then linearly unstable to a thermal
perturbation when $T>T_{\rm ign}$ at the specified column
density. This is a linear instability and needs no ``trigger''. It is
just that (on the unstable branch) any slight upward
temperature perturbation will lead to more rapid nuclear energy
release than cooling. There is a stable branch for the ignition
condition at higher temperatures (see heavy dashed line in Figure 1),
where a positive temperature perturbation leads to more rapid cooling.
The dark solid and dashed line in Figure 1 displays the
complete ignition relation (equation [\ref{eq:ignition}]) for a
neutron star of mass $1.4 M_\odot$ and $R=10 \ {\rm km}$ accreting
matter with $Y=0.3$. 

\begin{figure}
\centering{\epsfig{file=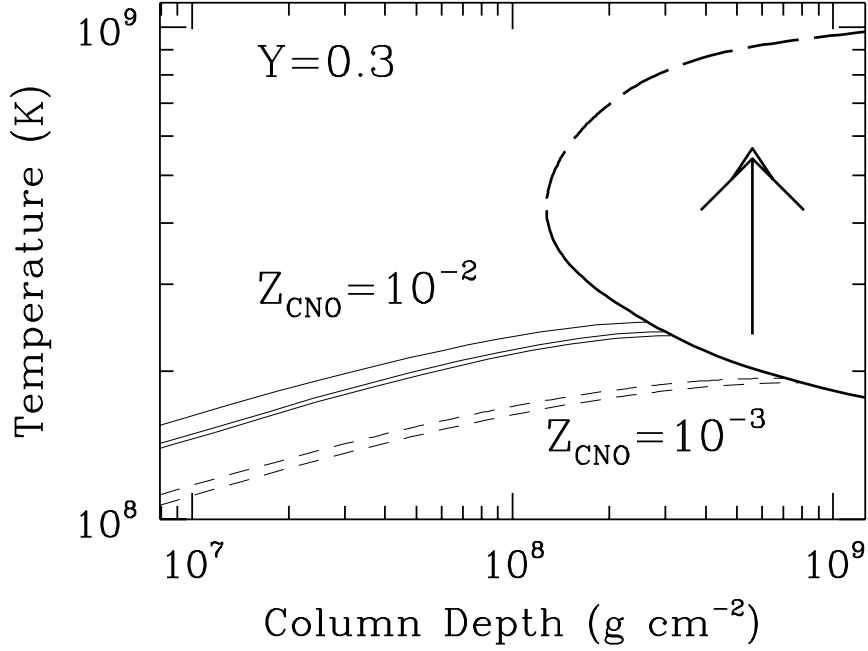,height=9cm}}
\caption{Ignition conditions on an accreting neutron star of mass
$M=1.4 M_\odot $ and $R=10\ {\rm km}$. The thick solid curve is the
unstable portion of helium ignition (equation [\protect\ref{eq:helign}]).  The
atmosphere is thermally unstable to the right of this line. The heavy
dashed line is the stable branch. The three thin solid lines are
settling solutions (from bottom to top) for $Z_{CNO}=10^{-2}$ and
$\dot m_5=0.3, \ 0.75$ and 3. The two thin dashed lines are for
$Z_{CNO}=10^{-3}$ and $\dot m_5=0.15$ and 0.75. The upward pointing
arrow shows the direction in which the temperature evolves when the
thermal instability occurs.}

\end{figure}
 Thermally unstable ignition of the accreting matter will occur if the
settling matter reaches the unstable part of the ignition curve (dark
solid) before burning all of the helium.  Now, what is the temperature
at the bottom of the accumulating matter?  The flux in the atmosphere
prior to the unstable helium ignition is just that generated from the
stable hydrogen burning, $F(y)=\epsilon_h(y_{t}-y)$, which when
combined with the radiative transfer equation gives a relation
\begin{equation}\label{eq:tyrel}
T^4={3\kappa \epsilon_h y_t^2\over 2 a c }. 
\end{equation}
It is the intersection of this solution with the ignition condition
(equation [\ref{eq:helign}] ) that tells us the temperature 
\begin{equation}
T_{\rm ign}\approx 3.5 \times 10^8 {\rm K} 
{\kappa^{1/18}\over Y^{1/6}}\left( Z_{CNO}\over \mu g_{14}\right)^{1/9},
\end{equation}
and column density 
\begin{equation}
y_{\rm ign}\approx
{2\times 10^7 \ {\rm g \ cm^{-2}}\over Y^{1/3} \kappa^{7/18} Z_{CNO}^{5/18}}
\left(1\over \mu g_{14}\right)^{2/9},
\end{equation}
for the unstable ignition of helium in the hydrogen-rich environment.
Both of these parameters are relatively insensitive to the value of
the opacity. It is important to point out that the ignition pressure,
$P=gy_{ign}\propto g^{7/9}$ and therefore the outcome of the unstable
helium flash depends on the neutron star mass and radius. This is in
contrast to thermonuclear instabilities on accreting white dwarfs,
where the ignition pressure is nearly independent of gravity, thus
allowing for simple relations between the recurrence time and
accretion rate (Shore, Livio and van den Heuvel 1994).

The electrons are mildly degenerate at the ignition site,
where the typical density is  $\rho\sim (5-8)\times 10^5 {\rm g \ cm^{-3}}$. 
Putting in characteristic values for these quantities, 
namely $Y=0.3$ and $\mu=0.6$ (we could correct for degeneracy to 
$\mu_{eff}$, but the ignition column is not that sensitive to it), 
we find that  the ignition conditions are 
\begin{equation}\label{eq:tign}
T_{\rm ign}\approx 
{2.12\times 10^8 \ {\rm K}}
\left(\kappa\over {0.04 \ {\rm cm^2 \ g^{-1}}}\right)^{1/18}
\left(Z_{CNO}\over 0.01\right)^{1/9}
\left(1.87\over g_{14}\right)^{1/9},
\end{equation} 
and 
\begin{equation}\label{eq:yign}
y_{\rm ign}\approx 
{3.7\times 10^8 \ {{\rm g}\over \ {\rm cm^{2}}}}
\left({0.04 \ {\rm cm^2 \ g^{-1}}}\over \kappa\right)^{7/18}
\left(0.01\over Z_{CNO}\right)^{5/18}
\left(1.87\over g_{14}\right)^{2/9},
\end{equation} 
where we have scaled the opacity to the values found from calculations
which include electron conduction. This formula compares well with
that of Fujimoto et al. (1987), who found $y_{ign}\propto
Z_{CNO}^{-0.5}$. These calculations also tell us that the physical
thickness of the accumulated layer prior to the instability is 10
meters, an important point for later reference. After discussing the
different accretion rate regimes for burning, we will come back and
use these results for predicting the nature of the resulting Type I
X-Ray bursts.

The three light solid lines are the envelope conditions just prior to
unstable helium ignition for $Z_{CNO}=10^{-2}$ and local accretion
rates of $\dot m_5=0.3, \ 0.75, $ and 3 (Bildsten \& Brown 1997).
These solutions should be viewed as upper limits to the ignition
column density, as they presume that there is no flux at the bottom
($y=y_t$) of the accumulating material. Note that, as derived above,
the ignition conditions are nearly independent of the local accretion
rate.  The ignition conditions found numerically are $y_{ign}\approx
3\times 10^8 {\rm g \ cm^{-2}}$, $T_{ign}\approx 2.4\times 10^8 {\rm
K}$ and $\rho\approx 8\times 10^5 \mdense$, a factor of 0.45 less
dense than we would have guessed if the electrons were non-degenerate
(see equations [\ref{eq:tign}] and [\ref{eq:yign}]). In this case,
degeneracy is starting to play a role. The flux exiting the atmosphere
is larger than just $\epsilon_h y_{ign}$ due to the internal
gravitational energy release. 

We also show two settling solutions for a lower metallicity of
$Z_{CNO}=10^{-3}$ and $\dot m_5=0.15$ and 0.75 (the two dashed lines).
For metallicities this low the ignition conditions become dependent on
$\dot m$ (see discussion in \S 3.3). This is evident in Figure 1,
where $y_{ign}\approx 7\times 10^8 \column$ for $\dot m_5=0.75$ and
$y_{ign}\approx 8\times 10^8 \column$ for $\dot m_5=0.15$.  Degeneracy
is becoming more important ($\mu_{eff}\approx 0.3 \mu$), thus changing
the opacity to $\kappa\approx 0.01 {\rm cm^2 \ g^{-1}}$. It is still
the case though that our equations (\ref{eq:yign}) and (\ref{eq:tign})
can reasonably well represent what is found in the numerical
simulations for the ignition.

\subsection { The Accretion Rate Dependence of Unstable Burning} 

We now ask how the nature of the unstable burning depends on the local
accretion rate. It ends up that there are two critical accretion rates
that happen to overlap with the range of accretion rates relevant to
the bright X-ray binaries. 

 At very high $\dot m$'s, the helium
unstably ignites at the bottom of the freshly accumulated pile prior
to hydrogen exhaustion, or $y_h> y_{ign}$ (Taam \& Picklum 1978, 1979,
Taam 1980). This translates into a critical accretion rate
\begin{equation}\label{eq:mdotc1} 
\dot m_{c1}\approx 
{{4.2\times 10^3 \ {\rm g \ cm^{-2}\ s^{-1}}}\over g_{14}^{2/9} X}
\left({0.04 \ {\rm cm^2 \ g^{-1}}}\over \kappa\right)^{7/18}
\left(Z_{CNO}\over 0.01 \right)^{13/18},
\end{equation} 
so that when $\dot m> \dot m_{c1}$ the helium ignites in a
hydrogen-rich environment. This formula agrees very well with the
results of Hanawa \& Fujimoto (1982) for $g_{14}<2$, even the scaling
with the surface gravity. It also reproduces the critical accretion
rates in Taam (1981) to within a factor of two for a range in
metallicities and neutron star masses and radii.\footnote{The
literature is littered with multiple designations for these critical
accretion rates that can lead to some confusion. We are following the
notation of the recent review article by Lewin et al. (1995). Just for
the sake of clarity, note that our $\dot M_{c1}$ is the same as $\dot
M_{st} (A)$ of Fujimoto et al. (1981) and $\dot M_{c2}$ is the same as
$\dot M_{st} (B)$. For some reason 1 and 2 are swapped in the
article by Taam (1981). }

For accretion rates below this value ($\dot m< \dot m_{c1}$), the
hydrogen burns steadily and accumulates a pure helium shell underneath
it. The column density of hydrogen on the star is just given by
equation (\ref{eq:yh}), and beneath this a pure helium shell
accumulates with time until a thermal instability is eventually
reached.  The helium ignition is always unstable. We will not derive
the necessary conditions for the helium flash, but rather find the
accretion rate below which the hydrogen burning shell becomes
thermally unstable. This will happen when the temperature in the
hydrogen burning region becomes low enough (roughly $T<8\times 10^7 \
{\rm K}$, see \S 2.3) so that the proton capture reactions again
determine the time to go around the CNO cycle loop. In this case, the
temperature sensitivity of the CNO burning ($\nu\approx 11/T_8^{1/3}$
from the $^{14}$N(p,$\gamma$)$^{15}$O reaction) leads to thermally
unstable hydrogen burning. The temperature at the bottom of the
hydrogen burning shell (where $\epsilon_h y=X E_h \dot m$) is
$T^4=3\kappa\epsilon_h y^2/2ac$, so that $T({H \ Shell})\approx 7\times
10^8 \ {\rm K} (\kappa X^2 \dot m_5/Z_{CNO})^{1/4}$. For stable
hydrogen burning, we must have $T({H\ Shell})> 8\times 10^7 \ {\rm K}$
or $\dot m> \dot m_{c2}$ where
\begin{equation}\label{eq:mdotc2} 
\dot m_{c2}\approx 
{{6.3\times 10^2 \ {\rm g \ cm^{-2}\ s^{-1}}}\over X}
\left({0.04 \ {\rm cm^2 \ g^{-1}}}\over \kappa\right)^{1/2}
\left(Z_{CNO}\over 0.01 \right)^{1/2}.
\end{equation}
For accretion rates $\dot m< \dot m_{c2}$ the hydrogen ignition is unstable 
and initiates a flash. Our formula for $\dot m_{c2}$ reproduces the
quoted numbers in Fujimoto et al. (1981), Taam (1981) and Wallace et al.
(1982) to within a factor of two for a large range in 
metallicities and neutron star masses and radii. This is 
reasonably good given the level of our approximations. 

The ratio of these two critical accretion rates, $\dot m_{c1}/\dot
m_{c2}\approx 5$ for a particular neutron star, pointing out that the
unstable ignition in a pure helium layer occurs only for a limited
range of accretion rates. These three critical accretion rates that
delineate at least four different states of time dependent burning are
summarized in Table 1.  Just for the ease of comparing to other
results, the global rates for a $1.4M_\odot$, $R=10 \ {\rm km}$
neutron star accreting matter with $Z_{CNO}=10^{-2}$ and $X=0.7$ are
also displayed. FHM also outlined the three time dependent cases,
calling them cases 1-3, which are also noted in Table 1.

\begin{table}[htb]
\begin{center}
\caption{Nuclear Burning Regimes at High Accretion Rates}
\begin{tabular}{ll}
\hline 
Range in Local Accretion Rate   & Type of Nuclear  Burning \\ 
\hline
$\dot m > \dot m_{st}$  & Stable hydrogen/helium burning in a \\
$\ \ \ \dot M>2.6\times 10^{-8} M_\odot \ {\rm yr^{-1}}$ &mixed
H/He environment  \\
& [Equations (\ref{eq:stabmdot}) and  (\ref{eq:mdots})]\\
 &\\
$\dot m_{st}> \dot m > \dot m_{c1}$ & Thermally unstable helium
ignition \\
$\ \ \ 2.6\times 10^{-8}>\dot M/(M_\odot {\rm yr^{-1}})
> 10^{-9} $  & in a mixed H/He environment \\
\ \ \  (FHM Case 1)  & [Equation (\ref{eq:mdotc1})]\\ 
 &\\
 $\dot m_{c1}> \dot m > \dot m_{c2}$ & Thermally unstable pure He ignition \\
$\ \ \ 10^{-9} > \dot M/(M_\odot {\rm yr^{-1}})> 2\times
10^{-10} $ 
& after complete hydrogen burning  \\
\ \ \  (FHM Case 2) & [Equation (\ref{eq:mdotc2})] \\ 
 &\\
$\dot m_{c2}> \dot m $ & Thermally unstable hydrogen burning \\
$\ \ \ 2\times 10^{-10} M_\odot {\rm yr^{-1}}> \dot M$ & triggers
combined flash\\ 
\ \ \ (FHM Case 3) & \\
\hline

\end{tabular}
\end{center}
\end{table}

\subsection{Helium Ignition at High Accretion Rates}

 We initially assumed that the heat flux coming through the envelope
prior to ignition was just that generated from the hot CNO cycle, or a
value $F_{CNO}\approx \epsilon_h y_{ign}$ just prior to ignition. The
typical ignition column density found in equation (\ref{eq:yign}) was
then independent of the local accretion rate and solely a function of
the metallicity and neutron star parameters. This approximation breaks
down at high accretion rates or low metallicities (possibly due to the
spallation scenario of Bildsten et al. 1992) where other sources of
energy release from the accumulating matter become important.  It is
easiest to see what happens by writing a flux (actually a combination
of gravitational energy release and deeper burning) $F_{deep}=E_{deep}
\dot m$ and ask what $E_{deep}$ needs to be so as to compete with the
flux in the envelope from the hot CNO cycle. This immediately tells us
that if $E_{deep}\simgreat (2\times 10^{17} {\rm ergs \ g^{-1}}/\dot
m_5) (Z_{CNO}/0.01)^{13/18}$, or only 0.2 MeV per accreted nucleon,
then we need to reconsider the thermal state of the accumulating
matter. We now go into this physics in some detail, as many of the
interesting results from {\it EXOSAT} and {\it RXTE} are for neutron
stars in this regime.

Let's start by rederiving the ignition condition for $F\propto \dot
m$. Presume that the flux flowing through the envelope prior to
ignition is $F=E_{deep} \dot m$ and leave $E_{deep}$ as a free
parameter.  Then combining the ignition condition (equation
[\ref{eq:helign}]) with the temperature profile for an atmosphere
carrying a constant flux $F$ gives
\begin{equation}\label{eq:yignhi}
y_{ign}(\dot m)\approx 
{1.8\times 10^8 {\rm g \ cm^{-2}}\over g_{14}^{4/13}}
\left({0.04 \ {\rm cm^2 \ g^{-1}}}\over \kappa\right)^{7/13}
\left(10^{18} {\rm ergs \ g^{-1}} 
\over E_{deep} \dot m_5\right)^{5/13}
\end{equation}
which we compare to the previous ignition column density
(equation [\ref{eq:yign}]) to find the condition on accretion rate,
metallicity and $E_{deep}$ that must be satisfied so that the
ignition column depends on $\dot m$. This gives 
\begin{equation}
E_{deep}\simgreat {10^{17} {\rm ergs \ g^{-1}}  \over \dot m_5}
\left({0.04 \ {\rm cm^2 \ g^{-1}}}\over \kappa\right)^{0.39}
\left(1\over g_{14}\right)^{0.222}
\left(Z_{CNO}\over 0.01\right)^{13/18}.
\end{equation}
So when this condition is satisfied we should expect that
the critical column density depends on the accretion rate in the
way given in equation (\ref{eq:yignhi}). 

Now, just how big is $E_{deep}$? Well, there are a few energy sources
to consider. The simplest is the energy released as matter falls a few
scale heights in the atmosphere (i. e. the heat lost due to the
non-adiabatic gravitational compression of the matter). This is
roughly $E_{deep}\approx 5 k_B T/2\mu m_p$ or $10^{17} \ {\rm ergs \
g^{-1}}$ at $T=3\times 10^8 \ {\rm K}$.  For normal metallicities,
this energy release would only be relevant at $\dot m$'s so high that
the burning is already stable. This energy release is critical at
lower metallicities and Bildsten \& Brown (1997) show this for a case
of $Z_{CNO}=10^{-4}$. Figure 1 shows a case for $Z_{CNO}=10^{-3}$,
where this energy release is included.  A much larger energy source is
the deeper burning of residual hydrogen left from a previous flash
(Ayasli \& Joss 1982, Wallace et. al. 1982, Taam et al. 1993).  Taam
et al. (1993, 1996) have noted that the residual hydrogen left from a
previous flash cannot burn until electron capture densities are
reached (this is presumably due to the slowness of proton captures on
the high Z nuclei). In this limit a fraction $X_r$ of the accreting
hydrogen is burned at large depths, giving $E_{deep}=7.7\times 10^{18}
\ {\rm ergs \ g^{-1}} X_r$. This energy release is so large that even
just a few percent of unburned hydrogen can be important.

Let's compare our simple estimates to the recent calculations of
successive thermal instabilities by Taam et al. (1996).  Their neutron
star had $M=1.4 M_\odot$, $R=9.1 \ {\rm km}$ and $X=0.7$. They
followed the time dependent evolution for models ranging from
$0.1-1.0$ times the Eddington limit. None of the models went into
steady-state burning, so for this case $\dot M_{st}$ is clearly in
excess of $\dot M_{\rm Edd}$. All models were time dependent and had
substantial residual hydrogen, ranging at the highest $\dot m$'s from
$X_r=0.08-0.21$.  Our formula gives $y_{ign}\approx 10^8 \ {\rm g \
cm^{-2}}$ over this range. Their accumulated columns as inferred from
the time between ignitions was nearly constant at $y\approx \dot m
t_{rec}\approx 5\times 10^7 \ {\rm g \ cm^{-2}}$, and so our estimate
is not so bad. 

The energy release from deeper regions leads to, on average, higher
ignition temperatures (up to $T=(3-4)\times 10^8 \ {\rm K}$) and lower
ignition pressures. This allows for the distinct possibility that
during the accumulation prior to ignition, even the Hot CNO cycle is
modified, as breakouts can occur at these temperatures via the
$^{15}$O($\alpha$, $\gamma$)$^{19}$Ne and $^{19}$Ne(p,
$\gamma$)$^{20}$Na reactions (see Champagne \& Wiescher 1992 for an
overview). The lower ignition pressures also leads to lower 
peak temperatures during the bursts and typically less fuel burned
in a particular burst (Taam et al. 1996). 

\section{Time Dependent Burning and Type I X-Ray Bursts}

What really happens as time marches on and accretion continuously
dumps matter onto the star? The simplest time-dependent behavior is a
limit cycle. The star would accumulate fuel for a time
$t_{rec}=y_{ign}/\dot m$ until the thermal instability is reached, at
which point the temperature rapidly rises and all of the fuel is
burned. The first time-dependent numerical calculations found this
type of behavior, but did not follow recurrent cycles.  All that was
found was the first burst and the ignition conditions for that burst
were comparable to what is derived here. Before discussing in some
detail what was learned from {\it EXOSAT} observations of Type I X-Ray
bursts, let's discuss a few of the numbers that can be reliably
predicted from theory.

\subsection{Burst Energies, Recurrence Times and Durations} 

When in the highest accretion rate unstable regime (see Table 1), the
ignition columns (see equations [\ref{eq:yign}] and [\ref{eq:yignhi}])
are in the range $(5-20)\times 10^7 \ {\rm g \ cm^{-2}}$, giving
recurrence times of $t_{rec}\sim 10^3 {\rm s}/\dot m_5$ or a few hours
between bursts for a neutron star accreting at $10^{-9} M_\odot \ {\rm
yr^{-1}}$.  The burst energy is $E_{burst}\approx 4\pi R^2
y_{ign}E_{nuc}$ if all of the fuel is burned, giving
\begin{equation}\label{eq:energy}
E_{burst}\approx 
5\times 10^{39} {\rm ergs}\left(R\over 10 \ {\rm km}\right)^2
\left(y_{ign}\over 10^8\  {\rm g\ cm^{-2}}\right)
\left(E_{nuc}\over 4\times 10^{18} \ {\rm ergs \ g^{-1}}\right).
\end{equation} 
The flux from the burst is close to the Eddington limit, so that the
duration is about the time it takes to radiate the nuclear energy at the
Eddington limit, $t_{burst}\approx E_{burst}/L_{Edd} \sim 10-20 \ {\rm
s}$. These three numbers, the burst energy, recurrence time and
duration, are all in agreement with the observations and are the
``hallmarks'' of our understanding that Type I X-Ray bursts arise from
thermonuclear instabilities in the accreted matter.

Let's briefly discuss what happens as the thermal instability develops
into a full-scale burst. There are many papers describing the time
dependent behavior of the burning (e.g. Taam \& Picklum 1979, Ayasli
\& Joss 1982, Wallace et al. 1982, Taam et al. 1993, 1996) and the
resulting thermonuclear reactions of the ``rp process'' which produce
elements beyond iron (Wallace \& Woosley 1981, Taam 1985, Van Wormer
et al. 1994, Schatz et al. 1997). As emphasized by FHM, the pressure
at the bottom of the accumulated pile is an important quantity for the
evolution of the flash. This is because the flash occurs at fixed
pressure, and eventually the matter becomes radiation dominated. For a
typical ignition column of $10^8 \column$, the pressure is
$P=gy\approx 10^{22} \ {\rm ergs \ cm^{-3}}$, so $aT_{max}^4/3\approx
P$ gives a a maximum temperature $T_{max}\approx 1.5\times 10^9 {\rm \
K}$. At this temperature, the degeneracy is lifted and the thermal
time (from equation [\ref{eq:thermal}]) at the bottom of the burning
material is about one second. In one-dimensional (i.e. spherically
symmetric) burning, convection nearly always occurs before these
maximum temperatures are reached. However, this only acts to transport
the heat upwards a few scale heights, as the convective zone never
reaches the photosphere (Joss 1977).  The energy is ultimately taken
out of the star via radiative diffusion.  The physical thickness of
the burning layer is $H\approx 5k_BT/2\mu m_pg\approx 18.5 \ {\rm m}
T_9$, or about 20 meters thicker than before the thermal instability
set in.

\subsection{What Was Learned from {\it EXOSAT} Observations}

The original information about Type I X-ray bursters came from
low-Earth orbit satellites. These discovered the phenomena and the
combination of theory and observation led to a firm belief that the
bursts arose from thermonuclear instabilities (see Lewin et. al. 1995). 
However, the frequent data gaps and absence of long-term
observations hindered studies of the accretion rate 
dependence of the phenomena. 

The 3.8 day orbit of {\it EXOSAT} was an excellent match for long-term
monitoring of bursters, thus revealing the dependence of their nuclear
burning behavior on accretion rate. As noted above, in the simplest
picture of a recurrent cycle, the time between bursts should decrease
as $\dot M$ increases since it takes less time to accumulate the
critical amount of fuel. Exactly the {\it opposite} behavior was
observed from many low accretion rate ($\dot M\sim 10^{-9} M_\odot \
{\rm yr^{-1}}$) neutron stars. A particularly good example is 4U
1705-44, where the recurrence time increased by a factor of $\approx
4$ when the accretion rate increased by a factor of $\approx 2$
(Langmeier et al. 1987, Gottwald et al. 1989). This is difficult to
understand in terms of what we discussed in \S 3. If the star is
accreting matter with $Z_{CNO}=10^{-2}$ then these accretion rates are
at the boundary ($\dot M\sim \dot M_{c1} $) of unstable helium
ignition in a hydrogen-rich environment at high $\dot M$ and unstable
pure helium ignition at lower $\dot M$. The expected change in burst
behavior as $\dot M$ increases would then be to more energetic and
more frequent bursts. This was not observed. The difficulties are
still present if we reduce the metallicities via spallation.

Other low $\dot M$  neutron stars showed similar behavior (see
Bildsten's [1995] discussion of many of the ``atoll'' sources), each
of which showed longer (and less periodic) recurrence times as $\dot
M$ increased.  More importantly, the amount of fuel consumed in the
bursts became much less than that accreted between bursts as $\dot M$
increased.  van Paradijs, Penninx \& Lewin (1988) tabulated this
effect for many X-ray burst sources and concluded that increasing
amounts of accreting fuel are consumed in a less visible way than Type
I X-ray bursts as $\dot M$ increases. The objects they studied have
all had Eddington limited radius expansion bursts and so the ratio of
the persistent flux to the flux during radius expansion measured the
accretion rate in units of Eddington accretion rate. From this, one
infers that these objects typically accrete at rates $\dot M \approx
(3-30)\times 10^{-10} \ M_\odot {\rm \ yr^{-1}}$. 

There are six higher accretion rate ``Z'' sources (Sco X-1, Cyg X-2,
GX 5-1, GX 17+2, GX 340+0, GX 349+2, Hasinger \& van der Klis 1989),
which are thought to be accreting near the Eddington limit of
$10^{-8}\ M_\odot \ {\rm yr^{-1}} $ (Fortner, Lamb \& Miller 1989,
Miller \& Lamb 1992). Their nuclear burning might be stable some of
the time, as they vary in accretion rate by a factor of $2-3$. None of
these are known to exhibit periodic bursting behavior, and only two
(Cygnus X-2 and GX 17+2) have ever shown Type I X-ray bursts. The
first burst found was from Cygnus X-2 (Kahn \& Grindlay 1984). It had
a rise time $<2.56 \ {\rm s}$ and a decay time $\approx 5-10 \ {\rm
s}$, but little else was certain from this observation. Kuulkers, van
der Klis \& van Paradijs (1995) found nine events from Cygnus X-2
which are reasonably interpreted as Type I X-ray bursts. They all
lasted about 3 seconds, had peak fluxes less than the persistent flux,
and burst energies $\sim 10^{38} {\rm ergs}$. However, they were so
infrequent that the resulting $\alpha$ value was usually $>10^3$ and
sometimes $>10^4$.  In other words, these bursts were not burning all
of the accreted fuel. RXTE has seen one type I X-ray burst from Cygnus
X-2 (Smale 1997).  Many bursts have been observed from GX 17+2 (Tawara
et al. 1984, Sztajno et al. 1986, Kuulkers et al. 1997) all of which
have $\alpha\simgreat 10^3$ and long durations (typically 100 seconds)
implying that little of the accreted fuel is consumed in the bursts.

\section{Beyond One-Dimensional Theories} 

If Type I X-ray bursts are the only manifestation of unstable burning,
then frequent energetic bursts should be seen from all weakly magnetic
and sub-Eddington neutron stars. {\it EXOSAT} did not find see this
occur, nor have other satellites. One solution to this mystery is to
assume that the accumulating matter only covers 10 \% of the stellar
area (say in the equatorial region if the matter is fed in via a thin
accretion disk), thus increasing the local accretion rate. This might
help with the Atoll sources, as the burning would become stable when
the global rate exceeds $\dot M\approx 10^{-9} M_\odot {\rm
yr^{-1}}$. However, it implies that the fuel stays in the equatorial
region until it reaches ignition pressures. It is not clear that this
is possible, especially since the accumulated matter is substantially
lighter than the ashes from burning. This solution would not work for
the Z sources, as they have shown bursts (though infrequent) and would
be absolutely stable if the local accretion rate was a factor of 10
larger. One could appeal to the spherical flows and magnetic funneling
implied by some for the QPO phenomenology (Alpar \& Shaham 1985,
Miller \& Lamb 1992) to have factors of two enhancement in the local
accretion rate. This would allow some Z source to be stable and others
not.

Rather than invoking alternative accretion scenarios to explain the
burst phenomenology, let's presume that the accumulated matter covers
most of the star before it undergoes a thermonuclear instability and
see if there is some physics we have previously omitted that now needs
to be considered. The most likely candidate is the propagation of
burning fronts around the star.

\subsection{Ignition and Propagation of Burning in Bursts}

Once it was understood that the X-ray bursts were thermonuclear
flashes, a few workers in the field (Joss 1978, Ruderman 1981, Shara
1982, Livio \& Bath 1982) pondered the possibility that the burning
was not spherically symmetric. In this case, there would be an
ignition site somewhere on the star (most likely where $\dot m$ is
highest) that would start a nuclear ``fire'' which spreads around the
star by igniting the accumulated fuel. The bursts rise to a peak
luminosity in about one second and are known (from the spectral
evolution and energetics) to cover the whole star, thus requiring a
lateral propagation speed of at least $10^6 \ {\rm cm \ s^{-1}}$.  The
fundamental reason to seriously consider this scenario is that it
takes many hours to accumulate a thermally unstable pile of fuel, but
only 10 seconds to burn it. Thus, the conditions must be identical to
better than a part in 1000 for the local thermal instability to occur
simultaneously over the whole star. This seems difficult to arrange in
the accretion environment.

One of the best quotes about what might be gained by relaxing
the spherically symmetric assumption was from Joss (1978):
\begin{quote}
{\it ``Our computations do not reproduce the complex and
individualized burst structure and recurrence patterns that have been
observed in some burst sources.  It is possible that many of these
complexities result from violations of spherical symmetry due to
stellar rotation or non-spherical accretion. Such effects could allow a
limited portion of the neutron star surface to undergo a flash that
then ignites other regions, in a manner that varies from burst to
burst and from one source to another.''}
\end{quote}
Following this, a few  (Fryxell \& Woosley 1982, Nozakura, Ikeuchi
\& Fujimoto 1984, Bildsten 1993, 1995) have estimated the
lateral propagation speeds for burning fronts on neutron stars.  The
heating of the gas from the thermal instability occurs on a timescale
of seconds, which is $10^5-10^6$ times longer than the time it takes
for sound to cross a scale height. The atmosphere has plenty of time
to hydrostatically adjust to the changing temperature, thus making
detonations very unlikely. So, let's begin by estimating the lateral
deflagration speed.

  The cold fuel ahead of the moving burning front is ignited  by
diffusive heating from the ashes. A steady burning front thus has a
characteristic thickness, which is the distance where diffusive
heating is comparable to the nuclear heating $l\approx (acT^4_f/3
\kappa\rho^2\epsilon_{3\alpha})^{1/2},$ where $T_f$ is the
undetermined burning temperature. The front propagates through $l$ in
the time it takes the nuclear burning to heat the gas to $T_f$,
$t_{nuc}\approx C_p T_f/\epsilon_{3\alpha}$, giving a speed
\begin{equation}\label{eq:vdefl} 
v\sim {l\over
t_{nuc}}\sim \left( ac T_f^2\epsilon_{3\alpha}\over 3
C_p^2\rho^2\kappa\right)^{1/2}\sim {h\over
(t_{th}t_{nuc})^{1/2}}.
\end{equation} 
This estimate is sensitive to the value $T_f$, thus requiring a more
careful derivation than provided by (\ref{eq:vdefl}). However, it is a
good initial guess and gives (for $P\approx 10^{22} \ {\rm erg \
cm^{-3}}$) values of order $100-1000 \ {\rm cm \ s^{-1}}$. Higher
ignition pressures can increase this by a factor of $\sim 10$, but it
seems difficult (if not impossible) to reach the $10^6 \ {\rm cm \
s^{-1}}$ needed for bursts.

Though the exact value of the deflagration speed is uncertain, there
is an accurate lower bound that comes from considering the quenching
of a burning front.  Propagation is halted when the lateral thickness
of the front, $l$, exceeds a pressure scale height, $h=P/\rho g$. In
this case most of the released nuclear energy is radiated to vacuum
rather than used to heat adjacent unburned matter. Since $l\sim
h(t_{nuc}/t_{th})^{1/2}$, propagation requires enough fuel so that
$t_{nuc}\simless t_{th}$ at the burning temperature, translating into
a minimum column density (or pressure) for successful propagation (a
``flammability'' limit, which for pure helium accretion is
$y_{min}\approx 10^8 \ {\rm g \ cm^{-2}}$) and a minimum speed
$v_{min}\sim h/t_{th}\sim 100 \ {\rm cm \ s^{-1}}$ (Bildsten 1995).
\footnote{There is a possibility that the ignition columns become so
low at high $\dot M$'s that the burning fronts are completely
quenched, in which case no lateral propagation can occur. The
resulting time dependent burning in this limit would be very difficult
to observe, as the number of independent patches on the star would be
very large. This has yet to be calculated for the mixed
hydrogen/helium burning.}

The propagation speed is most likely much larger when convection is
occurring (Fryxell \& Woosley 1982).  The enhanced mixing and lateral
motions will increase the propagation speeds over the laminar values.
The convection is very efficient at transporting heat, as the time for
sound to cross a scale height $t_c\sim c_s/g\sim 10^{-6} \ {\rm s}$ is
much shorter than the local thermal time. Steady-state convection then
brings the temperature-density profile in the envelope very close to
the adiabatic value, regardless of the choice of a mixing length.
Crude dimensional analysis with a mixing length, $l_m$ gives a
convective velocity
\begin{equation}
v_c\sim
c_s\left(l_m\over h\right)^{1/3}\left(c_s/g\over t_{th}\right)^{1/3}\sim 10^6 \ {\rm cm \ s^{-1}},
\end{equation}
a value comparable to what is needed for the rise time of a Type I
X-ray burst.  The fact that the turbulent scale is close to a scale
height prohibits us from doing much more than saying that the
propagation speed is comparable to the convective velocity when
convection occurs.

\subsection{Lateral Propagation and Observations}

 Now, how do we expect the propagation speeds to affect the
observations?  In the presence of convection, the high combustion
speed ($v\simgreat 10^6 \ {\rm cm \ s^{-1}}$) leads to rapid ignition
($\simless 10 \ {\rm s}$) of all connected parts of the surface which
have enough fuel for convective combustion. This is the dominant
burning mode at low $\dot M$'s, where the column densities prior to
ignition are large. Most of the star is then convectively combustible
at any one time, so that a recurrent thermal instability leads to
quasi-periodic Type I X-ray bursting.

However, as $\dot M$ increases (especially in the range above $10^{-9}
M_\odot {\rm yr^{-1}}$) the ignition columns are decreasing (see \S
3.3 and Taam et al. 1996) and convection becomes less prevalent. In
this regime, the slower burning speeds are more prevalent. This breaks
the burst periodicity, reduces the amount of fuel available for Type I
X-ray bursts, and leads to slow burning of the accumulated matter on
the neutron star. In this regime, the combustion front moves at a
speed $\sim 1000 \ {\rm cm \ s^{-1}}$, so that a ``ring of fire'' of
horizontal width $vt_{th}\sim 100$ meters moves out from the local
instability. Most importantly, the time for the fire to go around the
star ($R/v\sim 10^3 {\rm s}$) is comparable to the fuel accumulation
time so that only a few fires are burning at once (Bildsten 1995).  A
single fire gives a luminosity profile that roughly looks like
\begin{equation}\label{eq:lumfire}
L_{fire}(t) \approx F (2\pi v t)(v t_{th})\approx 
E_{nuc} y_{ign} 2\pi v^2 t,
\end{equation}
where $F$ is the local flux. When only a few fires are burning at one
time, the nuclear burning luminosity appears as $\sim$ hour-long
flares of amplitude $L_n\approx E_{nuc}\dot M$, or $1-5$ \% of the
accretion luminosity (Bildsten 1995).

\subsection{Does Unstable Nuclear Burning Cause the Very Low Frequency Noise?}

Hasinger \& van der Klis's (1989) analysis of {\it EXOSAT}
observations of the brightest LMXBs characterized the time dependence
of the X-ray flux from many neutron stars. They found that excess
power in the mHz-Hz range was often present and sometimes correlated
with other source properties. This noise has a spectral density
between flicker noise ($\propto 1/f$) and even steeper ($\propto
1/f^2$) noise and was dubbed the ``very low frequency noise'' (VLFN)
by van der Klis et al. (1987). In the time domain, it looks like $\sim
1-5 \%$ luminosity variations on minute to hour long time-scales,
comparable to what we just described might be produced at high
accretion rates from the slow burning fires.

The best way to decide if the VLFN has anything to do with nuclear
burning is to see if the strength of the hour-long flares (i.e. the
VLFN) correlates with the Type I X-ray bursts. One would not expect a
correlation if the VLFN was from accretion fluctuations. Bildsten
(1995) collated the {\it EXOSAT } X-ray burst data with the measured
VLFN power and found a continuum of behavior in the brightest
accreting X-ray sources. On one end of the continuum are the sources
that always show strong VLFN and never show enough Type I X-ray bursts
to burn all of the fuel (the Z sources), while on the other end are
those objects which are always bursting and never show VLFN (the
lowest accretion rate Atoll sources).  The most illuminating objects
are three (4U 1705-44, 4U 1636-53 and 4U 1735-44) which show X-ray
bursts and VLFN simultaneously. The varying $\dot M$'s for these
objects allows them to display a continuum of behavior. The VLFN is
absent, or much reduced, when the neutron star is undergoing regular
Type I X-ray bursts, and increases in strength as the bursts become
less fuel-consuming and aperiodic at higher $\dot M$'s.

The correlation found by Bildsten (1995) points to the concept that as
the accretion rate increases, there is a tradeoff of nuclear fuel
between X-ray bursts and VLFN, favoring VLFN at high accretion
rates. It also implies that the burning should by asymmetric on the
star, allowing for measurements of the spin period. We discuss this
in \S 7.

\section{ The Role of a  Magnetic Field }

  A Type I X-ray burst has never been seen from a magnetized
($B\simgreat 10^{12} \gauss$) accreting pulsar.  This has led to the
oft-quoted quip that ``X-Ray pulsars don't burst''.  The lack of
bursts from accreting pulsars was surprising since they accrete at
rates comparable to the X-ray bursters.  Joss \& Li (1980) first explained
the lack of bursts by {\em stabilizing} the nuclear burning with a
high local accretion rate ($\dot m> \dot m_{st}$).\footnote{ Joss \&
Li (1980) also suggested that the reduction of the opacity in an
ultra-strong ($B>10^{13} \ {\rm G}$) field could stabilize burning for
$\dot m< \dot m_{st}$.  This is at odds with the recent abstract by
Lamb, Miller and Taam (1996), who suggested that the reduced opacity
would increase the critical accretion rate above which the burning
becomes stable. Our equation (\ref{eq:stabmdot}) gave $\dot m_{st}
\propto \kappa^{-3/4}$, and we agree with Lamb et al.'s suggestion.}
The magnetic field funnels the accretion onto the polar cap and
confines the accreted matter all the way to the ignition pressure. The
constraint of $\dot m> \dot m_{st}$ is easily satisfied for pulsars
with $\dot{M} \simgreat 10^{-10} M_\odot \yr^{-1}$, as the fractional
area of the polar cap only needs to satisfy $A_{\rm cap}/4\pi R^2
\simless 0.01$. This is well within the estimates obtained by either
following the field lines from the magnetospheric radius to the star
(Lamb, Pethick, \& Pines 1973), or allowing the matter to penetrate
through the magnetopause via a Rayleigh-Taylor instability and attach
to field lines at smaller radii (Arons \& Lea 1976; Elsner \& Lamb
1977).

However, given the range in magnetic field strengths and accretion
rates, it seems unlikely that this could explain the 
absence of Type I X-ray bursts in {\em all} accreting pulsars.  
Bildsten (1995) thus suggested that, even when the burning is 
thermally unstable, a strong magnetic field might inhibit the rapid
lateral convective motion needed for the combustion front to 
rapidly ignite the whole star (or polar cap).  The burning
front would then propagate at the slower speed set by diffusive
heat transport. In this case, the nuclear burning would appear as a flare
with a duration of a few minutes to an hour (depending on the 
propagation speed and 
the lateral extent of the fuel-rich region). These flares would 
rise on a timescale comparable to the duration and not necessarily
be asymmetrical in time. 

The field strength required to halt the convective ($\sim 10^6
\cm\secd^{-1}$) propagation of burning fronts is not known. Convection
is potentially stabilized when $B^2 > 8\pi P_{ign}$ (Gough \& Tayler
1966), which requires $B\simgreat 7\times 10^{11}\gauss$ in the helium
burning regions.  This is satisfied for most accreting
pulsars. However, there are subtleties even at weaker magnetic fields,
as the convection occurs in a region with $\rho\approx 8\times 10^5
{\rm \ g \ cm^{-3}}$, giving a local ram pressure when in steady-state
convection of $P_{ram}\approx \rho v_c^2\ll P $. This can most likely
push around a magnetic field with strength $B\simless (8\pi\rho
v_c^2)^{1/2}\approx \ 5 \times 10^9 {\rm G}$, but we do not really
know how convection (and most importantly burning front propagation)
is altered by magnetic fields in the interesting intermediate range
$\rho v_c^2\ll B^2/8\pi \ll P$ ($5\times 10^9 \ {\rm G} < B < 7\times
10^{11} \ {\rm G}$).

  We emphasize that the appearance of the instability (i.e., whether
it looks like a Type I X-ray burst or a flare lasting a few minutes)
will yield crucial information on the neutron star's surface magnetic
field. The accreting pulsars most likely to yield time-dependent
burning phenomena are either ones with low fields (like GRO J1744-28,
Bildsten \& Brown 1997) or very low accretion rates. Strohmayer et
al. (1997) searched for Type I X-ray bursts from GRO J1744-28, but
could not find any. The Type I X-ray bursts that they did detect were
from another source.  Perhaps the best objects to survey would be the
faint and weakly magnetic ``6-second pulsars'' (1E 2259+586, 1E
1048.1-5937 and 4U 0142+61, see Mereghetti \& Stella 1995). The
inferred fields from their steady spin down are $\simless 10^{12}\
{\rm G}$ and the global accretion rates are $\simless 10^{-11} M_\odot
\ {\rm yr^{-1}}$.  These objects could very well be unstably
burning. The burst (or flare) recurrence time and energetics
completely depends on whether the matter stays confined to the polar
cap prior to ignition. Magnetohydrostatic calculations (Hameury
et. al. 1983, Brown \& Bildsten 1997) suggest that the matter does
stay confined on the cap, in which case the recurrence time will be
shorter than expected for spherical accretion and the burst energetics
much smaller. Actually, the burst energy would just be
$E_{burst}\approx A_{cap}E_{nuc} y_{ign}$, a value much smaller
(potentially two to three orders of magnitude) than the typical Type I
X-ray burst. Gotthelf \& Kulkarni (1997) might have just detected such
an event from an unknown (and very faint, $L_Q<10^{33} \ {\rm ergs \
s^{-1}}$) object in the globular cluster M28.  They found a very
subluminous Type I X-ray burst with $L\sim 4\times 10^{36} {\rm ergs \
s^{-1}}$ and total burst energy of $2-3\times 10^{37} \ {\rm ergs}$,
consistent with complete burning over a small fraction of the star.

\section{Observed Periodicities During Type I X-Ray Bursts}
 
There were many early (but statistically weak) indications of
periodicities during Type I X-ray bursts that led Livio \& Bath (1982)
to speculate upon the possibilities of rotation or surface waves as
the source of periodicity. McDermott \& Taam (1987) calculated the
g-mode frequencies of the envelope during a type I X-ray burst and
showed that the modes would rapidly change their frequency (by a
factor of 2 in 10 seconds). The first convincing case of a highly
coherent oscillation during a Type I X-ray burst was that found by
Schoelkopf \& Kelley (1991) in Aql X-1 at 7.6 Hz. They argued that the
periodicity was the neutron star rotation period and that the signal
arose when the burning asymmetries were rotated in and out of view.

  Jongert \& van der Klis (1996) searched for pulsations from 147 Type
I X-Ray bursts detected by EXOSAT. Nothing was found, with typical
fractional RMS amplitude limits of 2-6 \% for frequencies up to
(typically) 256 Hz. The recent launch of the {\it Rossi X-Ray Timing
Explorer} (RXTE) has dramatically changed this situation. Even though
it is likely that our discussion of these results will be dated by the
time this article appears, it is important to summarize what has been
found so far. Table 2 contains a summary of what was known as of May
1997.

\begin{table}[htb]
\begin{center}
\caption{Rapid Periodicities During Type I X-Ray Bursts (as of
May 1997)}
\begin{tabular}{lllll}
\hline 
Object Name & $\nu_B$ (Hz)& $\nu_{d}$ (Hz) & & Reference\\
\hline
4U 1728-34 & 363 & 363 & &  Strohmayer et. al. 1996 \\
KS 1731-260 & 524& $260\pm 10$&  &  Smith et al. 1997, Wijnands \& van
 der Klis \\
Aql X-1 & 7.6,549 & -- &  & Schoelkopf \& Kelley 1991, Zhang et al. 1997b\\
4U 1636-53 & 581 & $276\pm 10$  &  &Zhang et al. 1997a, Wijnands et al. 1997 \\
MXB 1743-29 & 589 & -- & & Strohmayer et al. 1997\\
\hline
\end{tabular}
\end{center}
\end{table}

  Strohmayer et al. (1996) found nearly coherent $\nu_B=363$ Hz
oscillations during type I X-ray bursts from the atoll source 4U
1728-34. Pulsations with amplitudes of $2.5-10 \%$ were detected in
six of the eight bursts analyzed at that time. In addition, two
separate high frequency QPO's were found in the persistent
emission. These changed with accretion rate, but maintained a fixed
difference frequency of $\nu_d=363 $ Hz, identical to the period seen
during the bursts. They suggested a ``beat'' frequency-like model and
associated the 363 Hz feature with the neutron star spin.

Smith, Morgan \& Bradt (1997) found a $\nu_B=523.92 \pm 0.05$ Hz
oscillation during the peak of a single Type I X-ray burst from KS
1731-260. This burst had radius expansion and the periodicity was only
found at the end of the contraction phase. The oscillation was very
coherent ($Q>900$) and had a pulse fraction of 6\%. No harmonics were
detected. Wijnands \& van der Klis (1997) found kHz QPO's from this
object at a fixed difference frequency $\nu_d=260\pm 10$ Hz,
consistent with one-half the frequency seen during the bursts. Another
object, 4U 1636-53, has a difference frequency in the kHz QPO's of
$\nu_d=276\pm 10 $ Hz (Wijnands et al. 1997) and a burst frequency of
$\nu_B=581$ Hz (Zhang et al. 1997a), not quite consistent with twice
the difference frequency. 
Strohmayer et al. detected 589 Hz oscillations during three Type I
X-ray bursts from an object (most likely MXB 1743-29) near the
galactic center. These bursts were temporally separated by just over a
month and the oscillations were detected in the few seconds after the
burst peak. 

  The frequency during the bursts from both MXB 1743-29 and 4U 1728-34
increased by a few Hz as the neutron star atmosphere cools. There is
no torque large enough to change the neutron star spin this rapidly,
so Strohmayer et al. (1997) have argued that the observed periodicity
is the rotation rate of the burning shell. They noted that the slight
($\Delta r \ll R$) hydrostatic expansion during the burning can
explain the observations if the shell conserves angular momentum. In
that case, the change in thickness (estimated from \S's 3.1 and 4.1 to
be $\Delta r \approx 20$ meters) will lead to a frequency change of
the burning material by an amount $\Delta \nu\approx \nu (\Delta
r/R)\approx 2\times 10^{-3} \nu $, or a change of 1 Hz for a 500 Hz
rotation. This is close to what is observed. It is presumed that the
neutron star spin frequency is the higher value and is, of course,
unchanging throughout the burst.

 However, the observed frequency evolution is always from low to high
during the cooling phase of the burst. Why is this?  It is most likely
because the atmospheric scale height is decreasing during the cooling
phase. The atmosphere expanded and spun-down at the onset of the
thermonuclear instability. However, this spin-down is difficult to
observe, as the radiative layer (see discussion in \S 4.1) delays the
information about the burst, so that by the time the observer sees it,
the expansion and spin-down has already occurred. There might be some
bursts where the spin-down evolution can be detected. Such a detection
would obviously make us much more confident of Strohmayer et al's
(1997) interpretation.

  Their conjecture implies that the burning matter wraps around the
star a few times during the instability, as the relative shear is a
few Hz for many seconds. Shear layers are typically unstable to the
Kelvin-Helmholtz instability. However, the buoyancy due to the mean
molecular weight contrast or even thermal (i.e. the buoyancy in
equation [\ref{eq:brunt}]) can easily stabilize this shear (at least
for a dynamical time) as the Richardson number ${\it
Ri}=N^2/(dv_{rot}/dz)^2\gg 1/4$ (Fujimoto 1988). The shear might thus
persist for the few seconds required. Further theoretical studies need
to be carried out to fully understand the repercussions of this
result. Problems that immediately come to mind are the role of any
magnetic field and more importantly, the possibility of longer
timescale rotational instabilities (such as the baroclinic
instability). Basically, one must rule out any mechanism that can
transport angular momentum to the shell in less than ten seconds.
That is a very long time on the neutron star, about $10^7$ sound
crossing times!

\section{Acknowledgements} 

I would like to thank the organizers of this Advanced Study Institute
for carrying out a flawless meeting and Jan van Paradijs for his
patience with my late submission. This article has greatly benefited
from countless conversations with Ed Brown about the subtleties of
nuclear burning and settling on accreting neutron stars.  Andrew
Cumming simulated the signals expected from propagating combustion
fronts on rotating neutron stars and helped me think through the
theoretical repercussions of the recent RXTE results.  Tod Strohmayer
was kind enough to show me a large amount of RXTE data on
periodicities during Type I X-ray bursts and Bob Rutledge provided
early insights into the phenomenology of periodicities during the
bursts. This research was supported by NASA via grants NAG 5-2819 and
NAGW-4517 and by the Alfred P. Sloan Foundation.

\end{document}